\documentstyle[12pt,psfig]{article}

\begin{document}
{\tiny .}
\vspace{1.5cm}
\begin{center}
{\Large {\bf A Kiloparsec--Scale Hyper--Starburst in a 
Quasar  \\[3mm] Host Less than 1 Gigayear after the Big Bang}}\\[0.5cm] {\bf Fabian
Walter$^{*}$, Dominik Riechers$^{1*\#}$, Pierre Cox$^{\&}$, Roberto
Neri$^{\&}$, Chris Carilli$^-$, Frank Bertoldi$^+$, Axel Weiss$^{++}$,
Roberto Maiolino$^{**}$}
\end{center}

\noindent
{\small $^*$ Max--Planck--Institut f\"ur Astronomie, K\"onigstuhl 17, D--69117
Heidelberg, Germany \\ $^\#$ California Institute of Technology, 1200
E. California Blvd., Pasadena, CA, 91125, USA \\ $^{\&}$ Institut de
Radio Astronomie Millim\'etrique, 300 Rue de la Piscine, F--38406
St--Martin--d'H\`eres, France\\ $^-$ National Radio Astronomy
Observatory, P.O. Box O, Socorro, NM 87801, USA \\ $^+$ Argelander
Insitut f\"ur Astronomie, Auf dem H\"ugel 71, D--53121 Bonn, Germany
\\ $^{++}$ Max--Planck--Institut f\"ur Radioastronomie, Auf dem
H\"ugel 69, D--53121 Bonn, Germany \\ $^{**}$ L'Istituto Nazionale di
Astrofisica, Osservatorio Astronomico di Roma, I--00040 Monte Porzio
Catone, Roma, Italy\\ $^1$ Hubble Fellow\\}

{\bf The host galaxy of the quasar SDSS J114816.64+525150.3 (at
  redshift z=6.42, when the Universe was $<$1 billion years old) has
  an infrared luminosity of 2.2$\times$10$^{13}$ L$_\odot^{~1,2}$,
  presumably significantly powered by a massive burst of star
  formation$^{3,4,5,6}$.  In local examples of extremely luminous
  galaxies such as Arp~220, the burst of star formation is
  concentrated in the relatively small central region of $<100$\,pc
  radius$^{7,8}$. It is unknown on which scales stars are forming in
  active galaxies in the early Universe, which are likely undergoing
  their initial burst of star formation. We do know that at some early
  point structures comparable to the spheroidal bulge of the Milky Way
  must have formed.  Here we report a spatially resolved image of
  [CII] emission of the host galaxy of J114816.64+525150.3 that
  demonstrates that its star forming gas is distributed over a radius
  of $\sim 750$\,pc around the centre. The surface density of the star
  formation rate averaged over this region is $\sim$1000
  M$_\odot$\,year$^{-1}$\,kpc$^{-2}$.  This surface density is
  comparable to the peak in Arp~220, though $\sim$2 orders of
  magnitudes larger in area. This vigorous star forming event will
  likely give rise to a massive spheroidal component in this system.}

The forbidden $^2$P$_{3/2}\rightarrow^2$P$_{1/2}$ fine--structure line
of ionized Carbon ([CII]) at 158 microns provides effective cooling in
regions where atomic transitions cannot be excited, and therefore
helps gas clouds to contract and form stars. [CII] emission is thus
known to be a fundamental diagnostic tool of the starforming
interstellar medium$^{9,10}$. Given the very bright continuum emission
of the central accreting black hole of quasars in optical and
near--infrared wavebands, standard star formation tracers (such as
hydrogen recombination lines) cannot be used to study star formation
in these systems. The [CII] line is however much brighter than the
underlying far--infrared (FIR) continuum, thus making it a prime
choice to characterize star formation in quasar host galaxies.

We used the IRAM Plateau de Bure interferometer to resolve the [CII]
emission from the z=6.42 host galaxy of J114816.64+525150.3 (one of
the most distant quasars known$^{11,12}$; hereafter: J1148+5251) with
a linear resolution of $\sim1.5$\,kpc.  J1148+5251 is one of only two
sources for which the detection of [CII] emission is reported at high
redshift to date$^{5,13}$. A large reservoir of molecular gas
(2\,$\times$\,10$^{10}$\,M$_\odot$), the prerequisite for star
formation, has been characterized in this system through redshifted
rotational transition lines of carbon monoxide (CO)$^{3,4,14}$. At a
redshift of z=6.42 the age of the Universe was just $\sim870$ million
years (or 1/16th of its present age) and 1$"$ on the sky corresponds
to 5.6 kpc$^{15,16}$.

The distribution of the [CII] emission is shown in the middle panel of
Figure~1.  Gaussian fitting to the spatially resolved [CII] emission
gives an intrinsic source size of 0.27$"\pm$0.05$"$ (1.5$\pm$0.3 kpc).
The [CII] emission is embedded within the molecular gas reservoir
traced by CO$^{14}$, however the [CII] emission is offset to the north
from the optical quasar and the CO peak by $\sim$0.1$"$ ($\sim$600
pc). Given the good agreement between the position of the optical
quasar and the simultaneous 158 micron continuum observations (Fig.~1,
left) we do not attribute this offset to inaccurate astrometry.  The
significance of the [CII] detection is high enough that it shows
spatially resolved velocity structure (red and blue contours in the
right panel of Figure~1).

The (rest--frame) FIR continuum emission underlying the [CII] line is
detected at 10 sigma significance in the integrated frequency spectrum
(Fig.~2). If the FIR continuum was due to the (unresolved) optical
quasar, a 10 sigma point source is expected at the optical position.
However, from Figure~1 (left) we only find $\sim$50\% of the flux to
be coincident with the optical quasar position. This implies that the
sensitivity of our observations is not high enough to image the
remaining FIR flux that is presumably due to the more extended
emission from star formation. This would imply that at most 50\% of
the FIR emission can be attributed to heating by the central black
hole, i.e.  the FIR emission is significantly powered by star
formation (in good agreement with the molecular gas$^{3,4}$, dense
gas$^{17}$, radio continuum$^{6}$ and dust properties$^{1,2}$ of this
source). In the following we thus assume a FIR luminosity due to star
formation of $\sim$1.1\,$\times10^{13}$ L$_\odot$, i.e., a star
formation rate of $\sim$1700 M$_\odot$\,yr$^{-1}$ (assuming a standard
initial stellar mass function$^{1,18}$). The low significance of the
resolved FIR emission is the reason why it cannot be used to constrain
the size of the starburst region.

The compactness of the [CII] emission implies that massive star
formation is concentrated in the central region with radius 750\,pc of
the system, even though molecular material is available on larger
scales (but our [CII] observations cannot rule out star formation at
lower surface densities over the entire molecular gas reservoir).
Given the star formation rate derived above, we find an extreme
average star formation rate surface density of
$\sim$1000\,M$_\odot$\,yr$^{-1}$\,kpc$^{-2}$ ($\sim7\!\times\!10^{12}$
L$_\odot$\,pc$^{-2}$) over this central 750\,pc radius region.
Similarly high starburst surface densities are also found in the
centre of local ULIRGs such as Arp\,220 (where each nucleus of size
$\sim100$\,pc has L$_{\rm FIR}$=3$\times$10$^{11}$L$_\odot$), albeit
on spatial scales that are by two orders of magnitudes
smaller$^{7,8}$.  For comparison, the Galactic young starforming
cluster associated with Orion KL also exhibits such high densities in
its central region$^{20}$ (L$_{\rm FIR}$=1.2$\times10^5$\,L$_\odot$,
area: $\sim1$\,arcmin$^{2}$, 0.013\,pc$^2$, resulting in
$\sim10^{13}$\,L$_\odot$\,kpc$^{-2}$), however over an area that is 8
orders of magnitudes smaller than in J1148+5251.

In the context of other galaxies in the early universe, this
kpc--scale `hyper'--starburst has a star formation rate surface
density that is one order of magnitude higher than what is found in
massive starforming z$\sim$2.5 submillimeter galaxies$^{21}$. It is
however consistent with recent theoretical descriptions of (dust
opacity) Eddington limited star formation of a radiation
pressure--supported starburst on kpc scales$^{19}$. The high star
formation rate surface density is also compatible with other theories
describing `maximum starbursts'$^{22}$: stars can form at a rate
limited by SFR=$\epsilon\times M_{\rm gas}/t_{\rm dyn}$, where
$\epsilon$ is the star formation efficiency, $M_{\rm gas}$ is the gas
within radius $r$ and $t_{\rm dyn}$ is the dynamical (or free--fall)
time, given by $\sqrt {r^3/(2\,G\,M)}$. For $r$=750\,pc,
$M\!\sim\!M_{\rm gas}\sim$10$^{10}$\,M$_\odot$ a star formation
efficiency of $\epsilon\sim$0.4 is required to explain star formation
rate density that we observe in the case of J1148+5251. Such high
efficiencies may be expected given the high dense gas fractions found
in local ULIRGs$^{23}$. In this calculation, we assume that the
stellar initial mass function in this object is not significantly
different from what is known locally.  Such a high star formation
efficiency could be expected if J1148+5251 were to undergo a major
merger, where the gas is funneled to the central 1.5\,kpc on rapid
timescales. We note however that our observations do not provide clear
evidence for a merging system and that other mechanisms may be
responsible for fueling the ongoing starburst$^{24}$. We also note
that the star formation rate surface density of
$\sim$1000\,M$_\odot$\,yr$^{-1}$\,kpc$^{-2}$ is a value averaged over
the central $\sim$kpc, i.e. this value could be significantly higher
on smaller scales, which in turn may violate the theoretical
descriptions of `maximum starbursts'.

Our observations provide direct evidence for strong, kpc--scale star
formation episodes at the end of Cosmic Reionization that enable the
growth of stellar bulges in quasar host galaxies. Such `hyper
starbursts' appear to have an order of magnitude higher star formation
rate surface densities on kpc scales than previously studied systems
at high redshift$^{21}$. The observations presented here are currently
the best means by which to quantify star formation rates and their
surface densities in quasars at the earliest cosmic epochs. They thus
demonstrate that [CII] observations will play a key role in studies of
resolved star formation regions in the first Gyr of the Universe using
the upcoming Atacama Large Millimeter/submillimeter
Array (ALMA)$^{25}$.\\

%\newpage

\noindent
{\bf References:}\\

\noindent
1. Bertoldi, F., Carilli, C.L., Cox, P., Fan, X., Strauss, M.A.,
Beelen, A., Omont, A., Zylka, R., Dust and Molecular Emission from
High-redshift Quasars, {\em Astron. Astroph.} {\bf 406}, 55-58 (2003)

\noindent
2.  Beelen, A., Cox,P., Benford, D.J., Dowell, C.D, Kovacs, A.,
Bertoldi, F., Omont, A., Carilli, C.L. 350 micron Dust Emission from
High-Redshift Quasars,  {\em Astroph. J.} {\bf 642}, 694-701 (2006)

\noindent
3.  Walter, F., et al. Molecular gas in the host galaxy of a quasar at
redshift z = 6.42, {\em Nature} {\bf 424}, 406-408 (2003)

\noindent
4.  Bertoldi, F., et al. High-excitation CO in a quasar host galaxy at
z =6.42, {\em Astron. Astroph. Letter} {\bf 409}, 47-50 (2003)

\noindent
5. Maiolino, R., et al., First detection of [CII]158 $\mu$m at high
redshift: vigorous star formation in the early universe, {\em
Astron. Astroph. Letters} {\bf 440}, 51-54 (2005)

\noindent
6. Carilli, C., et al., Radio Continuum Imaging of
Far-Infrared-Luminous QSOs at z$>$6, {\em Astron. J.} {\bf 128},
997-1001 (2004)

\noindent
7. Downes, D., \& Solomon, P., Rotating Nuclear Rings and Extreme
Starbursts in Ultraluminous Galaxies, {\em Astroph. J.} {\bf 507},
615-654 (1999)

\noindent
8. Scoville, N.Z., Yun, M.S., Bryant, P.M., 'Arcsecond Imaging of CO
Emission in the Nucleus of Arp 220', {\em Astroph. J.} {\bf 484}, 702-719
(1997)

\noindent
9.  Tielens, A.G.G.M., \& Hollenbach, D. Photodissociation regions. I
- Basic model. II - A model for the Orion photodissociation region,
{\em Astroph. J.} {\bf 291}, 722-754 (1985)

\noindent
10.  Stacey, G.J., Geis, N., Genzel, R., Lugten, J.B., Poglitsch, A.,
Sternberg, A., Townes, C.H. The 158 micron forbidden C II line - A
measure of global star formation activity in galaxies, {\em
Astroph. J.} {\bf 373}, 423-444 (1991)

\noindent
11. Fan, X., et al., A Survey of z$>$5.7 Quasars in the Sloan Digital
Sky Survey. II. Discovery of Three Additional Quasars at z$>$6, {\em
Astron. J.} {\bf 125}, 1649-1659 (2003)

\noindent
12. Fan, X. et al., Constraining the Evolution of the Ionizing
Background and the Epoch of Reionization with z$\sim$6 Quasars. II. A
Sample of 19 Quasars, {\em Astron. J.} {\bf 132}, 117-136 (2006)

\noindent
13. Iono, D., Yun, M.S., Elvis, M., Peck, A.B., Ho, P.T.P., Wilner,
D.J., Hunter, T.R., Matsushita, S., Muller, S., A Detection of [C II]
Line Emission in the z = 4.7 QSO BR 1202-0725, {\em Astroph. J.
  Letter} {\bf 645}, 97-100 (2006)

\noindent
14. Walter, F., Carilli, C., Bertoldi, F., Menten, K., Cox, P., Lo,
K.Y., Fan, X., Strauss, M.A., Resolved Molecular Gas in a Quasar Host
Galaxy at Redshift z=6.42, {\em Astroph. J. Letter} {\bf 615}, 17-20
(2004)

\noindent
15. Spergel, D.N., et al., Three-Year Wilkinson Microwave Anisotropy
Probe (WMAP) Observations: Implications for Cosmology, {\em Astroph.
  J. Suppl.} {\bf 170}, 377-408 (2007)

\noindent
16. Wright, E.L., A Cosmology Calculator for the World Wide Web, {\em
Publ. Astron. Soc. Pac.} {\bf 118}, 1711-1715 (2006)

\noindent
17. Riechers, D.A., Walter, F., Carilli, C., Bertoldi, F.,
Observations of Dense Molecular Gas in a Quasar Host Galaxy at z=6.42:
Further Evidence for a Nonlinear Dense Gas-Star Formation Relation at
Early Cosmic Times, {\em Astroph. J. Letters} {\bf 671}, 13-16 (2007)

\noindent
18. Omont, A., Cox, P., Bertoldi, F., McMahon, R. G., Carilli, C.,
Isaak, K. G., A 1.2 mm MAMBO/IRAM-30 m survey of dust emission from
the highest redshift PSS quasars, {\em Astron. Astroph.} {\bf 374},
371-381 (2001)

\noindent
19. Thompson, T., Quataert, E., Murrey, N., Radiation
Pressure-supported Starburst Disks and Active Galactic Nucleus
Fueling, {\em Astrophy. J.} {\bf 630}, 167-185 (2005)

\noindent
20. Werner, M.W., Gatley, I., Becklin, E.E., Harper, D.A.,
Loewenstein, R.F., Telesco, C.M., Thronson, H.A., One arc--minute
resolution maps of the Orion Nebula at 20, 50, and 100 microns, {\em
  Astroph. J.} {\bf 204}, 420-426 (1976)

\noindent
21. Tacconi, L., et al., High-Resolution Millimeter Imaging of
Submillimeter Galaxies, {\em Astroph. J.} {\bf 640}, 228-240 (2006)

\noindent
22. Elmegreen, B.G., Galactic Bulge Formation as a Maximum Intensity
Starburst, {\em Astroph. J.} {\bf 517}, 103-107 (1999)

\noindent
23. Gao, Y., Solomon, P.M., HCN Survey of Normal Spiral,
Infrared--Luminous, and Ultraluminous Galaxies, {\em Astroph. J.
  Suppl.} {\bf 152}, 63-80 (2004)

\noindent
24. Dekel, A., Birnboim, Y., Engel, G., Freundlich, J., Goerdt, T.,
Mumcuoglu, M., Neistein, E., Pichon, C., Teyssier, R., Zinger, E., The
Main Mode of Galaxy Formation: Early Massive Galaxies by Cold Streams
in Hot Haloes, {\em Nature}, under review (2008)
 
\noindent
25. Walter, F., Carilli, C., Detecting the most distant (z$>$7)
objects with ALMA, {\em Astroph. \& Space Science} {\bf 313}, 313-316
(2008)

\noindent
26. White, R.L., Becker, R.H., Fan, X., Strauss, M.A., Hubble Space
Telescope Advanced Camera for Surveys Observations of the z = 6.42
Quasar SDSS J1148+5251: A Leak in the Gunn-Peterson Trough, {\em
Astron. J.} {\bf 129}, 2102-2107 (2005)

\noindent
27. Solomon, P.M. \& Vanden Bout, P.A., Molecular Gas at High Redshift, {\em
Ann. Rev. Astron. Astroph.} {\bf 43}, 677-725 (2005)

\noindent
28.  Malhotra, S., et al. Infrared Space Observatory Measurements of
[C II] Line Variations in Galaxies, {\em Astroph. J.} {\bf 491}, 27-30
(1997)

\noindent
29. Luhman, M.L, Satyapal, S., Fiuscher, J. Wolfire, M.G., Cox, P.,
Lord, S.D., Smith, H.A., Stacey, G.J., Unger, S.J. Infrared Space
Observatory Measurements of a [C II] 158 Micron Line Deficit in
Ultraluminous Infrared Galaxies, {\em Astroph. J. Letters} {\bf 504},
11-15 (1998)

\vspace{1cm}

{\bf Acknowledgements:} This work is based on observations carried out
with the IRAM Plateau de Bure Interferometer. IRAM is supported by MPG
(Germany), INSU/CNRS (France) and IGN (Spain). DR acknowledges support
from from NASA through a Hubble Fellowship awarded by the Space
Telescope Science Institute, which is operated by the Association of
Universities for Research in Astronomy, Inc., for NASA. CC
acknowledges support from the Max--Planck Gesellschaft and the
Alexander von Humboldt Stiftung through the
Max--Planck--Forschungspreis 2005.  FW and DR appreciate the
hospitality of the Aspen Center for Physics, where this manuscript
was written.\\

{\bf Competing Interest Statement:} The authors declare that they have
no competing financial interests.\\

{\bf Correspondence and requests for material} should be addressed to
F.W. (walter@mpia.de)

\newpage
\vspace{5cm}
\begin{figure}[t!]
   \begin{center} \hspace{0cm} \hspace{-1cm}
   \psfig{figure=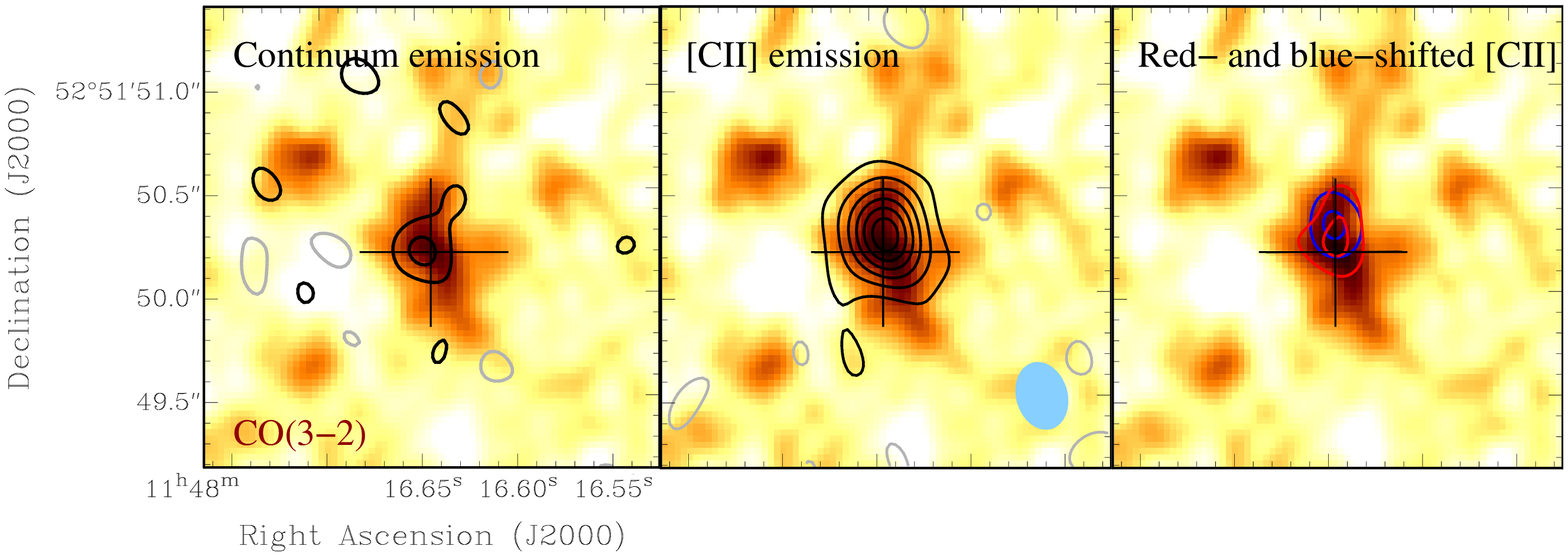,width=160mm,angle=0}
   \end{center} 
\vspace{-0.8cm}
\end{figure}

\noindent {\bf Figure 1:} [CII] observations of the z=6.42 quasar
J1148+5251 obtained with the IRAM Plateau de Bure interferometer.
Observations were obtained in the most extended antenna configuration
during three tracks in early 2007 and 2008 ($\nu_{\rm
  obs}$=256.17\,GHz, $\nu_{\rm rest}$=1900.54\,GHz), resulting in a
resolution of 0.31$''\times$0.23$''$ (1.7\,kpc$\times$1.3\,kpc; the
beamsize is shown in light blue colour in the middle panel).  The
resolved CO emission from VLA observations$^{14}$ is displayed as
colour scale in all three panels. The cross indicates the absolute
position (uncertainty: 0.03$''$) of the (unresolved) optical quasar as
derived from Hubble Space Telescope observations$^{26}$.  {\em Left:}
Contours represent the far--infrared continuum emission obtained from
the line--free channels of the [CII] observations integrated over a
445\,km\,s$^{-1}$ bandwidth (contour levels are --0.9 (grey), 0.9 and
1.8\,mJy (black); rms noise: 0.45\,mJy).  There is good agreement
between the optical quasar and the peak of the continuum emission, as
well as the peak of the molecular gas emission traced by CO,
demonstrating that our astrometry is accurate on scales of $<0.1''$.
{\em Middle:} Contours show the [CII] emission over a velocity range
of --293 to +293\,km\,s$^{-1}$ (contours are plotted in steps of
0.72\,mJy; rms noise: 0.36\,mJy).  The (rest--frame) beam--averaged
peak brightness temperature of the [CII] emission is 9.4$\pm$0.9\,K
(from the peak flux of 7.0$\pm$0.36\,mJy at a resolution of
0.31$''\times0.23''$), which is similar to the CO brightness
temperature (8.3\,K)$^{14}$. If the intrinsic temperature of the gas
were similar to that of the dust$^{2}$ (30--50\,K), this would imply
that we have not fully resolved the CO or the [CII] emission.  {\em
  Right:} Contours of blue-- and a red--shifted emission (averaged
over velocities from 75--175\,km\,s$^{-1}$ on either side) are plotted
as blue and red contours at 3.2 and 4.8\,mJy, respectively (rms noise:
0.63\,mJy). The dynamical mass of $\sim 10^{10}$\,M$_\odot$ within the
central 1.5\,kpc deduced from these observations (assuming v$_{\rm
  rot}\sim250$\,km\,s$^{-1}$) is in agreement with earlier estimates
on larger spatial scales$^{14}$.

\newpage

\begin{figure}[t!]
   \begin{center}
   \hspace{1cm}
   \psfig{figure=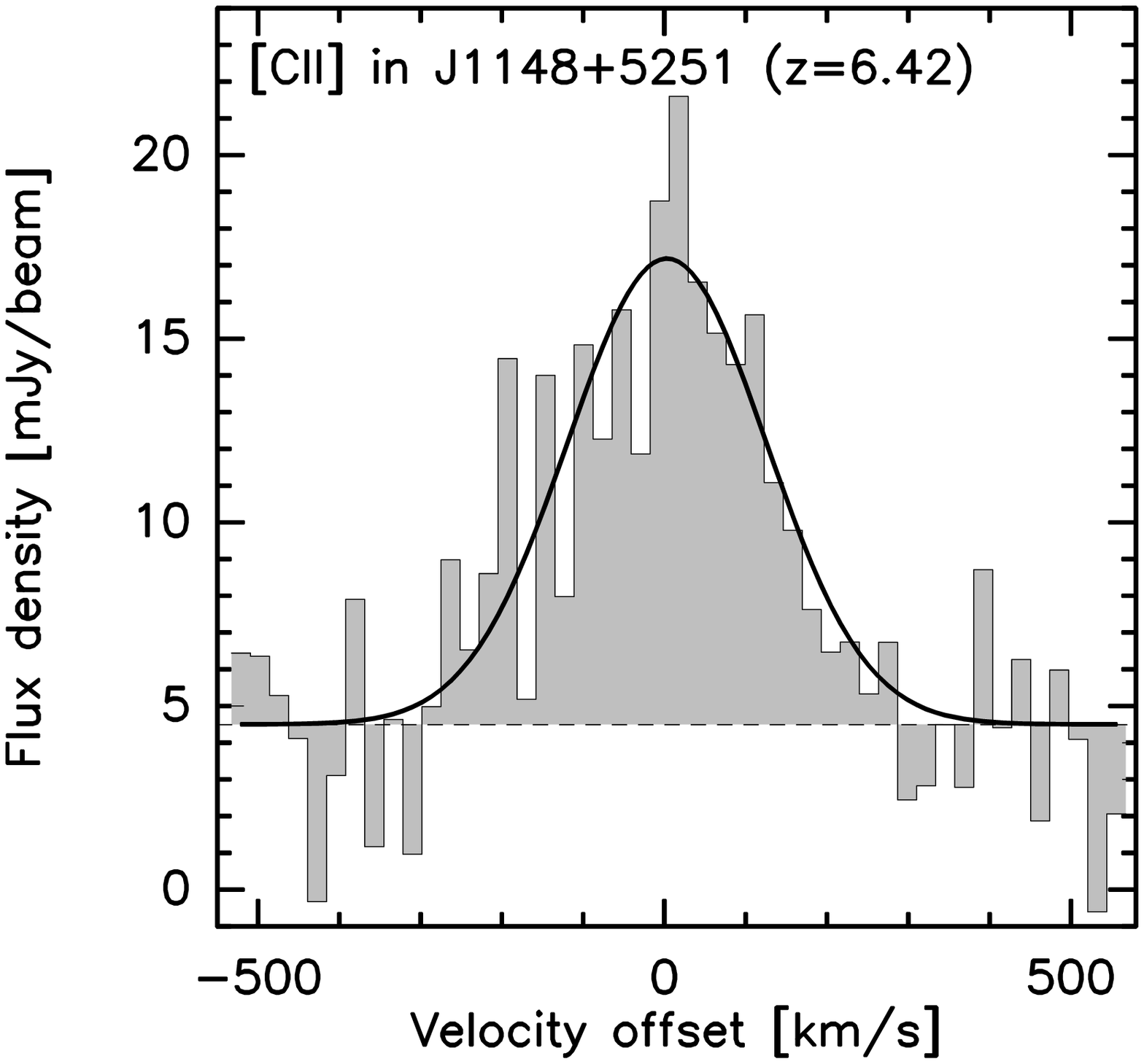,width=8cm,angle=-90}
   \end{center} 
%\vspace{-0.8cm}
\end{figure}

%\caption
\noindent {\bf Figure~2:} Spatially integrated [CII] spectrum of the
z=6.42 quasar J1148+5251. The [CII] line is detected at high
significance (bandwidth covered: 1\,GHz, or 1100\,km\,s$^{-1}$) and is
present on top of a 4.5$\pm$0.62\,mJy continuum (consistent with an
earlier estimate$^{1}$ of 5.0$\pm$0.6\,mJy).  Gaussian fitting to the
line gives a [CII] peak flux of 12.7$\pm$1.05\,mJy, a full width at
half maximum (FWHM) velocity of 287$\pm$28\,km\,s$^{-1}$ and a central
velocity of 3$\pm$12\,km\,s$^{-1}$ relative to the CO redshift$^{4}$
of z=6.419 ($\nu_{\rm obs}$=256.17\,GHz). This leads to a [CII] flux
of 3.9$\pm$0.3\,Jy\,km\,s$^{-1}$ (consistent with earlier, unresolved
observations$^{5}$ of 4.1$\pm$0.5\,Jy\,km\,s$^{-1}$), which
corresponds to a [CII] luminosity$^{27}$ of
L$'_{\rm[CII]}$=1.90$\pm$0.16$\times$10$^{10}$
K\,km\,s$^{-1}$\,pc$^{-2}$ or L$_{\rm
  [CII]}$=4.18$\pm$0.35$\times10^{9}$\,L$_\odot$ (adopting a
luminosity distance of D$_{\rm L}$=64 Gpc$^{16}$), yielding L$_{\rm
  [CII]}$/L$_{\rm FIR}$=1.9$\times10^{-4}$. This ratio is by an order
of magnitude smaller than what is found in local starforming galaxies
(a finding consistent with local ultraluminous infrared galaxies,
ULIRGs$^{5,28,29}$).  The line--free channels of the [CII]
observations are used to construct a continuum image of J1148+5251 at
158 microns (rest wavelength) as shown in Fig.~1 (left).

\end{document}